\documentclass[reprint,amsmath,amssymb,aps]{revtex4-1}

\usepackage{graphicx}
\usepackage{dcolumn}
\usepackage{bm}
\usepackage{color}

\newcommand{\eq}{Equation~}

\newcommand{\Fig}{Figure~}

\newcommand{\fig}{Figure~}

\newcommand{\ie}{i.e.,}

\newcommand{\ee}{\mathrm{e}}

\newcommand{\nablab}{\bm{\nabla}}

\newcommand{\sliver}{\hspace{0.12em}}  
\newcommand{\nlb}{\protect\nolinebreak\sliver}

\newcommand{\yr}{\nlb{\rm yr}}
\newcommand{\Myr}{\nlb{\rm Myr}}

\newcommand{\km}{\nlb{\rm km}}

\newcommand{\cmmm}{\nlb{\rm cm$^{-3}$}}

\newcommand{\cmms}{\nlb{\rm cm$^2$s$^{-1}$}}

\newcommand{\Gauss}{\nlb{\rm G}}

\newcommand{\new}[1]{#1}

\begin{document}

\title{Three-dimensional simulations of the magnetic stress in a neutron star crust}
\author{T.~S.~Wood}
\email{Toby.Wood@newcastle.ac.uk}
\affiliation{School of Mathematics and Statistics, Newcastle University, Newcastle upon Tyne, NE1~7RU, United~Kingdom}

\author{R.~Hollerbach}
\affiliation{Department of Applied Mathematics, University of Leeds, Leeds, LS2~9JT, United~Kingdom}

\date{\today}

\begin{abstract}
We present the first fully self-consistent three-dimensional model of a neutron star's magnetic field,
generated by electric currents in the star's crust via the Hall effect.
We find that the global-scale field converges to a
\new{dipolar}
Hall-attractor state, as seen
in recent axisymmetric models,
but that small-scale features in the magnetic field survive even on much longer timescales.
These small-scale features propagate toward the dipole equator,
where the crustal electric currents organize themselves into a strong equatorial jet.
By calculating the distribution of magnetic stresses in the crust,
we predict that neutron stars with fields stronger than $10^{14}$\Gauss\
can still be subject to starquakes more than $10^5$\yr\ after their formation.
\end{abstract}


\maketitle

\section{Introduction}
Neutron stars are of interest not only for the exotic states of matter they contain,
but also for their magnetic fields, which are the strongest in the universe.
In the case of pulsars,
\new{the rotation of the magnetic field}
produces beams of \new{non-thermal} radiation
that can be detected thousands of light-years away.
The field can also affect the dynamics of the star itself, through the magnetic stress it exerts
on the star's solid outer layer, or ``crust''.
In neutron stars with especially strong magnetic fields (known as magnetars)
the magnetic stress
can be strong enough to fracture the crust,
producing a starquake \citep{Ruderman91c}.
This is believed to be the mechanism behind the gamma-ray flares and X-ray outbursts detected in these objects
\citep{ThompsonDuncan01}.

The magnetic fields of pulsars, inferred from spindown measurements, can be anywhere in the range $10^8$--$10^{15}$\Gauss.  However, the spindown rate depends only on the large-scale component of the field at the magnetic poles, so these measurements may underestimate the actual field strength in the star.
\new{This could explain why some neutron stars produce magnetar-like emissions yet have slow spin-down rates
\citep{Gavriil-etal08,Rea-etal10,Scholz-etal14}.}
There is also
evidence that at least some neutron stars have stronger magnetic features on smaller scales
\citep{Guver-etal11,Gotthelf-etal13,Tiengo-etal13,Geppert-etal13}.
To interpret these observations it is necessary to develop a self-consistent,
three-dimensional (3D) model of neutron star magnetic fields.

The external field of a neutron star is generated by electric currents flowing within its crust \new{and core}.
The ions in the crust form a rigid lattice,
and the currents \new{there} arise purely through the flow of electrons,
whose dynamics depend primarily on the Hall effect \citep{GoldreichReisenegger92};
this situation is commonly referred to as electron magneto-hydrodynamics (EMHD).
Recently there have been numerous studies of EMHD in
neutron star crusts
\citep{ShalybkovUrpin97,HollerbachRudiger04,PonsGeppert07,
PernaPons11,KojimaKisaka12,Vigano-etal12,Pons-etal13,GC14b,Marchant-etal14}.
However,
because of difficulties solving the EMHD equations computationally in 3D,
in all of these studies the magnetic field was assumed to be axisymmetric.
Although there have been 3D studies of EMHD in simplified periodic-box geometry
\citep{Biskamp-etal99,ChoLazarian04,WareingHollerbach10},
the results
cannot be used to predict the global-scale field morphology in a real neutron star.
Currently,
the only predictions regarding the global-scale field come from axisymmetric models.

Using one such axisymmetric model,
\citet{GC14b} found that an initially dipolar magnetic field
evolves towards a quasi-steady configuration that they called a ``Hall attractor''.
After $\sim10^5$\yr\ the electric currents in the crust are concentrated in a narrow jet around the dipole equator,
producing a strong belt of poloidal field.
However,
it is unknown whether
this axisymmetric attractor would be
stable to 3D perturbations in a real neutron star.

Here, we present for the first time global 3D numerical simulations of the magnetic field in a neutron star crust.
We use a pseudo-spectral code that
allows us to perform simulations that are not only fully 3D, but also higher resolution than any of those previously presented even in 2D.
We describe how the Hall attractor
is modified in 3D,
and discuss the magnitude and distribution of magnetic stresses within the crust.

\section{The Model}
\label{sec:model}
We work in a reference frame that corotates with the neutron star's crust.
Because the ions are fixed within the crust, in this frame
the electric current, $\mathbf{J}$, depends only on the electron fluid velocity, $\mathbf{v}$.
The current is also directly related to the magnetic field, $\mathbf{B}$, via Amp\`ere's Law,
and we have (in Gaussian cgs units)
\begin{align}
  \mathbf{J} = -\ee n\mathbf{v} = \frac{c}{4\pi}\nablab\times\mathbf{B},
\end{align}
where $n$ is the electron number density,
$\ee$ is the elementary charge,
and $c$ is the speed of light.
If the crust is either sufficiently cool,
or close to isothermal,
then the magnetic field is frozen to the electron fluid \citep{GoldreichReisenegger92},
and evolves according to the equation
\begin{align}
  \frac{\partial\mathbf{B}}{\partial t} &=
  \nablab\times\left(\frac{c\mathbf{B}}{4\pi\ee n}\times(\nablab\times\mathbf{B})\right)
  - \nablab\times(\eta\nablab\times\mathbf{B}),
  \label{eq:induction}
\end{align}
where $\eta$ is the magnetic diffusivity.
The two terms on the right-hand side of \eq(\ref{eq:induction}) represent the Hall effect
and Ohmic dissipation, respectively.

In this study we approximate the structure of the crust as fixed and spherically symmetric,
and so the electron density $n$ and magnetic diffusivity $\eta$ are fixed functions of radius $r$.
We do not therefore take account of any deformations in the crust arising from magnetic stresses.
However, we can use our results to calculate the magnetic stress and determine whether it would be large
enough to induce fractures in the crust.
We take the top and bottom of the crust to be the spherical surfaces $r=R$ and $r=0.9R$ respectively,
where $R=10$\km\ is the radius of a typical neutron star.
For $n(r)$ and $\eta(r)$ we adopt the same analytical profiles used by
\citet{GC14}.
Specifically,
\begin{align*}
  n = n_0\!\left(1+\frac{1 - r/R}{0.0463}\right)^{\!4}
  \; \mbox{and} \;
  \eta = \eta_0\!\left(1 + \frac{1 - r/R}{0.0463}\right)^{\!-8/3},
\end{align*}
where $n_0=2.5\times10^{34}$\cmmm\ and $\eta_0=4.0\times10^{-4}$\cmms\
are the values at $r=R$.
These profiles are only rough approximations to the (highly uncertain) profiles in real neutron stars,
but fortunately our results are not sensitive to the specific profiles used.

Finally, we must impose boundary conditions on the magnetic field at the top and bottom boundaries of the crust.
At the top we impose vacuum boundary conditions, \ie\ we match to a current-free field outside the star.
At the bottom we impose either the idealized boundary conditions used by \citet{GC14},
which are $B_r=0$ and $J_r=0$, or the more realistic boundary conditions used by \citet{HollerbachRudiger04},
which model the star's core as a type-I superconductor.
From here on we refer to these two sets of conditions as GC and HR respectively.
In order to implement the HR boundary conditions, it is convenient to make a small modification to the
electron density profile $n(r)$, as described by \citet{HollerbachRudiger04},
to make $1/n$ vanish at the bottom of the crust.
We therefore use a slightly modified density profile $\tilde{n}(r)$, defined as $1/\tilde{n}(r) = 1/n(r)-1/n(0.9R)$.

The relative importance of the Hall effect and Ohmic dissipation terms in \eq(\ref{eq:induction})
depends on the strength of the magnetic field, $B_0$ say,
and is quantified by the Hall parameter, $H \equiv cB_0/(4\pi\ee n_0\eta_0)$.
In a typical magnetar, with
$B_0 = 10^{14}$\Gauss,
we have $H \simeq 50$,
implying that the Hall effect dominates the dynamics of the magnetic field.
In that case we expect the field to evolve on the Hall timescale,
\begin{equation}
  t_{\rm Hall} \equiv \frac{4\pi\ee n_0R^2}{cB_0} \simeq 1.6\Myr,
  \label{eq:Hall_time}
\end{equation}
where we have assumed that the characteristic lengthscale for the magnetic field is $R=10$\km.
If the field has strong, small-scale features,
then these will evolve on a shorter timescale.

To solve the EMHD equation~(\ref{eq:induction})
we have adapted the 3D MHD code PARODY,
developed by \citet{Dormy-etal98} and \citet{Aubert-etal08}.
The code is pseudo-spectral, and uses spherical harmonic expansions in latitude and longitude,
and a discrete grid in radius, making it perfectly suited to solving problems in spherical-shell geometry.
The results that we present here have a resolution of 128 grid-points in radius,
and spherical harmonics up to degree $l=100$.
We have benchmarked the code against previously published axisymmetric results
\citep{HollerbachRudiger02,HollerbachRudiger04,GC14} and find excellent agreement
in all cases.

\section{Results}
\subsection{Robustness of the Hall attractor}
\begin{figure*}
  \centering%
  \includegraphics[width=\textwidth]{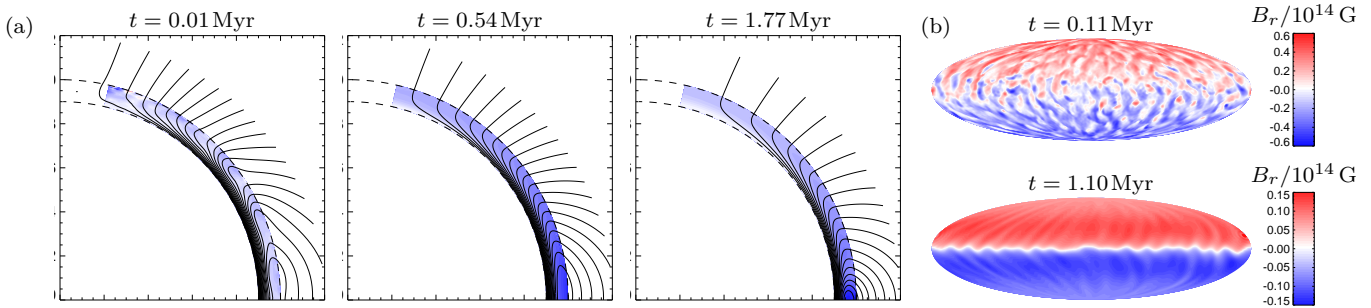}%
  \caption{(a) Lines of the azimuthally averaged poloidal field
    \new{and electron angular velocity}
    at successive times,
    illustrating convergence to the Hall attractor.
    The dashed lines indicate the boundaries of the crust.
    (b) The radial component of the magnetic field at the surface
    at early and late times in the same simulation.}%
    \label{fig:GC13_3D}%
  \end{figure*}

To determine whether the ``Hall attractor''
seen in earlier axisymmetric simulations is robust against 3D perturbations,
we have
repeated one simulation of \citet{GC14},
using the same boundary conditions and
dipolar
``Ohmic eigenmode''
initial condition for the axisymmetric component of the magnetic field.
To this initial axisymmetric poloidal field we add a low amplitude, small-scale 3D perturbation
with both poloidal and toroidal components,
\new{comprising spherical harmonics of degree $20 \leqslant l \leqslant 40$}.
As shown in \fig\ref{fig:GC13_3D}, the magnetic field
evolves towards a state broadly similar to the axisymmetric Hall attractor.
However, significant non-axisymmetric features persist in the simulation even on long timescales
(comparable to the global Hall timescale $t_{\rm Hall}$).
These features
would rapidly decay by the action of Ohmic dissipation alone,
so their longevity can be attributed to the Hall effect.
In fact it has previously been shown that, in the presence of a strong density gradient,
the Hall effect can sustain or even amplify small-scale features in the magnetic field \citep{Wood-etal14}.
\new{The length-scale of the longest-lived non-axisymmetric features is comparable to the largest scales
present in the initial 3D perturbations (degree $l=20$),
suggesting that they form by an up-scale transfer of magnetic energy.}

\new{The convergence of the global-scale magnetic field to the attractor state occurs on the Hall timescale~(\ref{eq:Hall_time}),
after which the evolution proceeds on the slower timescale of Ohmic diffusion.
This is in agreement with results from axisymmetric simulations
\citep{HollerbachRudiger04,GC14b}.
(In the simulations of \citet{Marchant-etal14},
by contrast,
the field was initialized already in a Hall equilibrium state,
and therefore evolved on the Ohmic timescale throughout.)
The poloidal magnetic flux becomes concentrated
into a ``belt'' around the equator of the magnetic dipole,
implying a strong equatorial
jet of electrons within the crust.
The angular velocity in this jet is approximately constant along each poloidal field line
(see \fig\ref{fig:GC13_3D}).
We do not find that the Hall effect enhances the diffusion of the magnetic field,
as originally suggested by \citet{GoldreichReisenegger92}.
However, by concentrating the poloidal flux around the equator, the Hall effect does
lead to a more rapid decline in the strength of the radial field at the poles.}

\begin{figure}
  \centering%
  \includegraphics[width=8.6cm]{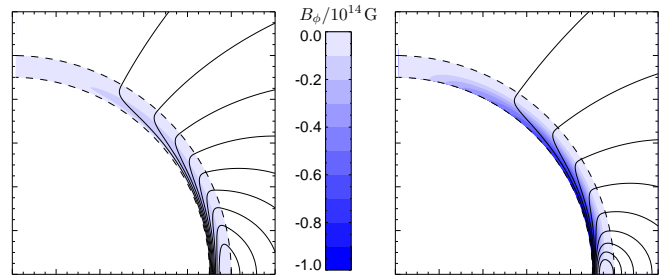}%
  \caption{Poloidal fieldlines and contours of toroidal field
  at $t = 1.2$\Myr.
  The left and right panels show the cases with GC and HR boundary conditions, respectively.}%
    \label{fig:BC-compare}%
  \end{figure}

\begin{figure*}
  \centering%
  \includegraphics[width=\textwidth]{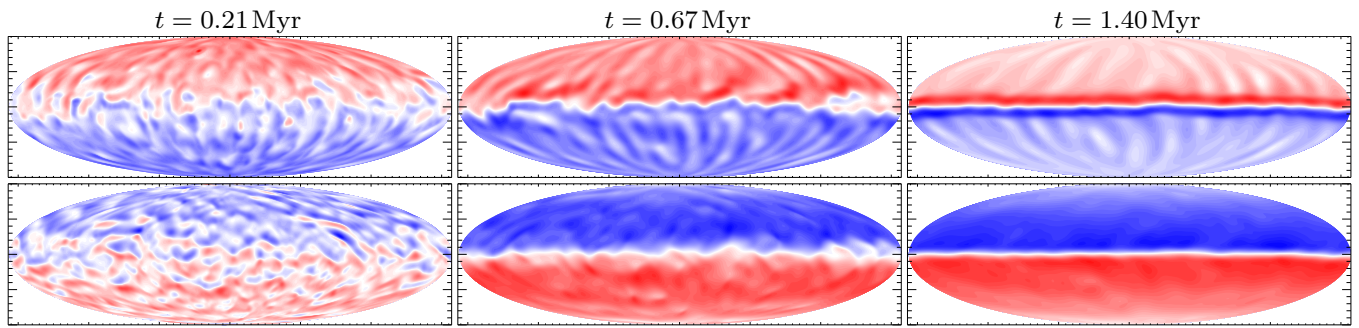}%
  \caption{Evolution of the magnetic field in the simulation with HR boundary conditions.
  Top row: $B_r$ at $r=R$.  Bottom row: $B_\phi$ at $r=0.9R$.
  Colorbars are adjusted to the maximum values in each plot, \new{and use a linear scale}.}%
    \label{fig:GC13_super}%
  \end{figure*}

The magnetic field is primarily poloidal, with a weaker toroidal component in the inner crust.
However, the GC boundary conditions used in this simulation
impose zero toroidal field
both at the top and bottom
of the crust.
In a real neutron star, currents
can flow
between the crust and core,
generating significant toroidal fields.
We have therefore repeated this simulation using the more realistic HR boundary conditions,
which allow for finite toroidal field at the lower boundary.
\Fig\ref{fig:BC-compare}
compares the azimuthally averaged magnetic field
at the same time in both simulations,
revealing that although the morphology of the poloidal field is modified,
a similar Hall attractor state still exists.  The toroidal field in the second simulation
is much stronger, as expected, but remains weaker than the poloidal field.
This toroidal field is associated with a meridional flow of electrons that is equatorward at the star's surface,
and drags the poloidal magnetic field lines.  As a result, the equatorial belt of poloidal flux,
and the corresponding electron jet, is even stronger in the simulation with HR boundary conditions,
and is
pushed slightly deeper into the crust.
\Fig\ref{fig:GC13_super}
illustrates the development of the equatorial belt,
which manifests as bands of positive and negative $B_r$
on either side of the equator at the star's surface.

  \begin{figure*}
  \centering%
  \includegraphics[width=\textwidth]{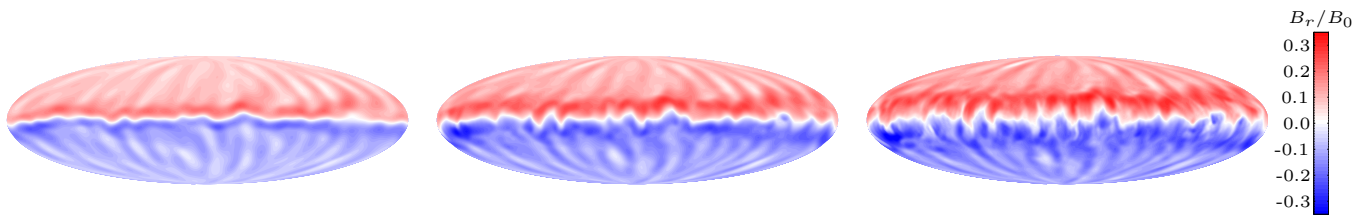}%
  \caption{The surface radial field in three simulations with
  $B_0=10^{14}$\Gauss, $2\times10^{14}$\Gauss, $4\times10^{14}$\Gauss\ (\ie\ $H=50,100,200$).
    }%
  \label{fig:GC13_strong}%
  \end{figure*}

\subsection{The effect of field strength}
Spin-down measurements of neutron stars tell us only the strength of the
large-scale magnetic field
near the poles.
However, it is clear from \fig\ref{fig:GC13_super} that the field strength elsewhere in the crust can greatly exceed
this observed value.
At the time plotted in \fig\ref{fig:BC-compare}b, for example, the strength of the surface magnetic field varies between
$\simeq5\times10^{12}$\Gauss\ at the dipole axis and $\simeq3\times10^{13}$\Gauss\ near the equator.
Near the bottom of the crust,
the field is stronger
by a further order of magnitude than at the surface.
The Hall effect may therefore be even more significant in neutron stars than previously thought,
and so we have repeated the same simulation shown in \fig\ref{fig:GC13_super}
with a stronger initial magnetic field (\ie\ a larger Hall parameter $H$).
Because the evolution of the global-scale magnetic field occurs on the Hall timescale~(\ref{eq:Hall_time}),
we expect that a stronger field will converge more rapidly to the Hall attractor,
before the small-scale magnetic features
have been dissipated by resistivity.
This is confirmed in \fig\ref{fig:GC13_strong},
which shows the surface radial field in three simulations with increasing magnetic field strengths.
In each case the field is plotted at time $t=0.6\,t_{\rm Hall}$,
which corresponds to $t \simeq 1$, 0.5, and 0.25\Myr\ respectively,
and an equatorial jet of electrons has already formed.  In the case with the strongest magnetic field,
the jet is broader and more spatially disordered; at later times in the same simulation, the jet becomes
increasingly laminar, but remains broader than in the other simulations.

\subsection{The crustal magnetic stress}
\begin{figure}
  \centering%
  \includegraphics[width=8.6cm]{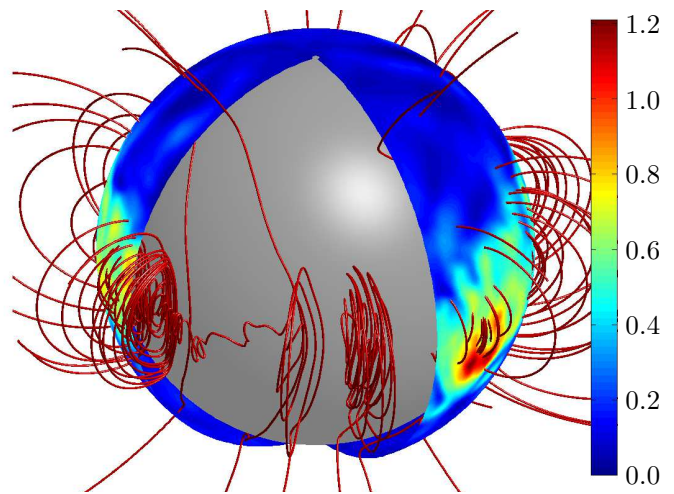}%
  \caption{Lines of the magnetic field generated by currents in the crust.
  The coloring indicates the ratio of magnetic pressure to breaking stress at the surface of the crust.
  A portion of the surface has been cut away to show fieldlines inside the crust, whose lower boundary
  is the gray sphere.}%
  \label{fig:stress}%
  \end{figure}

If the magnetic shear stress within the crust exceeds the breaking stress of the ionic lattice
then it will induce a crustal fracture.
Molecular dynamics models indicate that the breaking stress is approximately $5\times10^{-3}$ of the electron pressure, $P_\ee$ \citep{ChugunovHorowitz10}.
At each point within the crust, the strongest magnetic shear stress is exerted on a surface that makes an
angle of $45^\circ$ to the local magnetic field direction,
and this stress is exactly equal to the magnetic pressure, $P_{\rm m}$.
The condition for fracturing is therefore $P_{\rm m} \gtrsim 5\times10^{-3}P_\ee$,
where
$P_{\rm m} = |\mathbf{B}|^2/(8\pi)$ and $P_\ee \simeq (3\pi^2n)^{4/3}\hslash c/(12\pi^2)$.
Fractures are most likely near the surface of the crust, where the density and pressure are lowest,
and the energy released in a near-surface fracture can directly power flares and outbursts.
In \fig\ref{fig:stress} we plot the ratio of magnetic pressure to breaking stress at the surface
of the crust in the same simulation shown in the last panel of \fig\ref{fig:GC13_strong},
revealing that patches around the dipole equator are susceptible to crustal fracturing.
These patches are localized in both latitude and longitude,
as a consequence of the three-dimensionality of the magnetic field.
They persist on long timescales, of order $t_{\rm Hall}$, and
propagate azimuthally in the direction of the electron jet.

\section{Discussion}
Our results indicate that neutron starquakes are most common in the vicinity of the electron jet within the crust,
which typically forms a ring around the dipole axis.
This jet forms on the Hall timescale,
which is dependent on the overall strength of
the magnetic field.
At earlier times the structure of the magnetic field is dependent on the initial conditions for the
proto-neutron star, which unfortunately are not well known.
The surface field is strongest in localized patches
around the magnetic equator,
where the local field strength typically exceeds that at the poles by an order of magnitude.

In our simulations, surface magnetic features migrate equatorward as a result of the toroidal magnetic field in the crust
and the associated meridional flow of electrons.
This migration could potentially be reversed if
a strong toroidal field of the opposite sign were present,
corresponding to a poleward flow of electrons at the star's surface
\citep{GeppertVigano14}.
There is no obvious process that can generate such a strong toroidal field within the crust,
but it could be
a remnant from the star's formation \citep{ThompsonDuncan93}
or the result of toroidal flux expulsion from the core \citep{Lander14}.
To fully describe the processes that can impart a strong toroidal field to the crust,
it will be necessary to couple our crust model to a model of the superconducting core.

The flow of heat from a cooling neutron star is inhibited across magnetic field lines,
so the strong concentration of poloidal magnetic flux in the equatorial belt
could
trap heat within the electron jet
\citep{Vigano-etal13}.
The breaking stress is very sensitive to temperature
\citep{ChugunovHorowitz10}, and so the equatorial region of the crust may be even more susceptible to
fracturing than we have found here.  We are currently extending our model to describe the
thermal--magnetic evolution of the crust, and its coupling to the core.

\begin{acknowledgments}
T.S.W.~and R.H.~were supported by STFC grant ST/K000853/1.
Computations were performed on ARC1 and ARC2, part of the High Performance Computing facilities at the University of Leeds.
We thank
Konstantinos Gourgouliatos,
C\'eline Guervilly,
David Ian Jones,
Maxim Lyutikov,
and the anonymous referees
for helpful comments and suggestions.
\end{acknowledgments}

\end{document}